\documentclass[aip,graphicx]{revtex4-2}
\usepackage[dvipdfmx]{graphicx}
\usepackage{color}
\usepackage{mathrsfs}
\usepackage{amsmath,amssymb}

\def\u#1{_{\rm #1}}

\draft % marks overfull lines with a black rule on the right

\begin{document}

% Use  the \preprint command to place your local institutional report number 
\preprint{}

\title{Prediction of $^1$H singlet relaxation via intermolecular dipolar couplings using the molecular dynamics method}

\author {K. Miyanishi}
\email{miyanishi@qc.ee.es.osaka-u.ac.jp}
\affiliation{Division of Advanced Electronics and Optical Science, Department of Systems Innovation, Graduate School of Engineering Science, Osaka University, 1-3 Machikaneyama, Toyonaka, Osaka 560-8531, Japan}
\affiliation{Center for Quantum Information and Quantum Biology, Osaka University, 1-2 Machikaneyama, Toyonaka 560-8531, Japan}
\author{W. Mizukami}
% mail: wataru.mizukami.857@qiqb.otri.osaka-u.ac.jp
\affiliation{Center for Quantum Information and Quantum Biology, Osaka University, 1-2 Machikaneyama, Toyonaka 560-8531, Japan}
\affiliation{JST, PRESTO, Kawaguchi, Japan}
\author{M. Motoyama}
% mail address
\affiliation{Division of Advanced Electronics and Optical Science, Department of Systems Innovation, Graduate School of Engineering Science, Osaka University, 1-3 Machikaneyama, Toyonaka, Osaka 560-8531, Japan}
\author{N. Ichijo}
% mail: naoki.ichijo@murata.com
\affiliation{Division of Advanced Electronics and Optical Science, Department of Systems Innovation, Graduate School of Engineering Science, Osaka University, 1-3 Machikaneyama, Toyonaka, Osaka 560-8531, Japan}
\author{A. Kagawa}
% mail: kagawa@ee.es.osaka-u.ac.jp 
\affiliation{Division of Advanced Electronics and Optical Science, Department of Systems Innovation, Graduate School of Engineering Science, Osaka University, 1-3 Machikaneyama, Toyonaka, Osaka 560-8531, Japan}
\affiliation{Center for Quantum Information and Quantum Biology, Osaka University, 1-2 Machikaneyama, Toyonaka 560-8531, Japan}
\affiliation{JST, PRESTO, Kawaguchi, Japan}
\author{M. Negoro}
% mail: negoro@qiqb.osaka-u.ac.jp 
\affiliation{Center for Quantum Information and Quantum Biology, Osaka University, 1-2 Machikaneyama, Toyonaka 560-8531, Japan}
\affiliation{Institute for Quantum Life Science, National Institutes for Quantum and Radiological Science and Technology, 4-9-1, Anagawa, Inage-Ku, Chiba 263-8555, Japan}
\author{M. Kitagawa}
% mail: kitagawa@mail.osaka-u.ac.jp
\affiliation{Division of Advanced Electronics and Optical Science, Department of Systems Innovation, Graduate School of Engineering Science, Osaka University, 1-3 Machikaneyama, Toyonaka, Osaka 560-8531, Japan}
\affiliation{Center for Quantum Information and Quantum Biology, Osaka University, 1-2 Machikaneyama, Toyonaka 560-8531, Japan}

\date{\today}

\begin{abstract}
Dissolution dynamical nuclear polarization has been applied in various fields, including chemistry, biology, and medical science.
To expand the scope of these applications, the nuclear singlet state, which is decoherence-free against dipolar relaxation between spin pairs, has been studied experimentally, theoretically, and numerically.
% However, the synthesis of the new molecule and isotope substitutions for the long relaxation time takes time, effort, and a high cost.
The singlet state composed of proton spins is used in several applications, such as enhanced polarization preservation, molecular tag to probe slow dynamic processes, and detection of ligand--protein complexes. 
In this study, we predict the lifetimes of the nuclear spin states composed of proton spin pairs using the molecular dynamics method and quantum chemistry simulations.
We consider intramolecular and intermolecular dipolar, chemical shift anisotropy, and spin--rotation interactions.
In particular, the relaxation rate of intermolecular dipolar interactions is calculated using the molecular dynamics method for various solvents. 
The calculated values and the experimental values are of the same order of magnitude . Our program would provide insight into the molecular design of several NMR applications and would be helpful in predicting the nuclear spin relaxation time of synthetic molecules in advance. 
\end{abstract}


\maketitle %\maketitle must follow title, authors, abstract and \pacs

\section{Introduction}

% Proton nuclear magnetic resonance~(NMR) is widely used in various fields because the natural abundance ratio of $^1$H is almost unity, most of the molecules contain protons, and $^1$H has a high gyromagnetic ratio.
% Recently, a solvent containing hyperpolarized nuclear spins generated by 
Dissolution dynamic nuclear polarization~(dissolution DNP) has been developed to improve the sensitivity of NMR. It was applied to several nuclear magnetic resonance ~(NMR) and magnetic resonance imaging~(MRI) applications such as \textit{in vivo} metabolic imaging, drug discovery, and characterization of protein conformation~\cite{golman2006metabolic,Keshari2014,ARDENKJAERLARSEN20163,COMMENT201639,KIM2019501,Ragavan2017Real-Time_folding}.
% These experiments need to be performed within the longitudinal relaxation time $T_1$, which is the time to relax the hyperpolarized state to the thermally polarized state.
In these experiments, $^{13}$C NMR is relatively often used because $^{13}$C has a relatively long longitudinal relaxation time $T_1$, which is the time to relax a nuclear spin state to a thermally polarized state.
Therefore, hyperpolarized $^{13}$C spins can be useful as long-lived hypersensitive sensors.
$T_1$ of the proton is usually within ten seconds owing to its high gyromagnetic ratio, and this short relaxation time limits the application scope of hyperpolarized proton NMR.

A scheme using the nuclear singlet state, which is a spin state composed of two homonuclear spin-1/2 nuclei, was demonstrated to prolong the nuclear spin relaxation time~\cite{Carravetta04JACS,Carravetta04PRL,levitt2012singlet}.
The long relaxation time of the singlet state opens up new applications in NMR spectroscopy and magnetic resonance imaging ~(MRI)~\cite{salvi2012boosting,buratto2014drug,buratto2014exploring,buratto2016ligand,miyanishi2020long, cavadini2005slow,sarkar2007singlet,sarkar2008measurement,ahuja2009diffusion,pileio2015real,tourell2018singlet,mamone2018nuclear, warren2009increasing,vasos2009long,tayler2012direct,franzoni2012long,pileio2013recycling,theis2016direct,roy2016long}.
To achieve longer relaxation times, the synthesis of isotope-substituted molecules and a new design of a molecular unit have been proposed~\cite{stevanato2015nuclear,HillCousins2015,kiryutin2019proton}.
Isotope substitution has a problem in that the synthesis of a molecule takes much time, effort, and cost.
Therefore, it is advantageous to predict the NMR relaxation time of the synthesized molecule in advance.

The theory of calculation of the longitudinal relaxation rate ~(1/$T_1$=$R_1$) was well established by Abragam and Redfield~\cite{abragam1961principles,redfield1965theory,kowalewski2006nuclear}, and that of the singlet nuclear spin relaxation rate ~(1/$T_{\mathrm{S}}$=$R_{\mathrm{S}}$) was also well established~\cite{pileio2009theory,pileio2010relaxation,pileio2020relaxation}.
The relaxation mechanisms were investigated experimentally, theoretically, and numerically~\cite{abragam1961principles,pileio2009theory,pileio2010relaxation,pileio2011singletviaIDC,pileio2011singletviaSCoSK,pileio2012long,ghosh2012determination,stevanato2015nuclear,zhang2016limits,haakansson2017prediction,redfield1965theory,kowalewski2006nuclear,peter2001calculation,case2002molecular,calero20151h,singer2017molecular,wildenberg2020computational}.
In most cases, the dipolar interaction between the nearest homonuclear spins, $j$ and $k$~(DDjk), is the primary relaxation mechanism for longitudinal relaxation.
However, the singlet state composed of spins $j$ and $k$ is immune to DDjk interaction.
These predictions require the Hamiltonian of the spin system based on the quantum chemical (QC) calculation and the rotational and translational correlation time of the fluctuating Hamiltonian.
These correlation times can be estimated using the Stokes--Einstein--Debye relation, which is derived from a classical hydrodynamic perspective, and electrostatic interaction is not considered.

The spin orders of protons are more prone to relaxation mechanisms for the following two reasons. First, protons have a higher gyromagnetic ratio than other nuclear species ($^{13}$C, $^{15}$N, etc.). Second, the relaxation rates due to intramolecular and intermolecular dipolar interactions become larger than those of other nuclear species. 
However, singlet spins composed of proton spins are used in several applications, such as long-term preservation of enhanced polarization~\cite{vasos2009long,ahuja2010proton,kiryutin2012creating,eills2021singlet},
molecular tag to probe slow dynamic processes~\cite{sarkar2007singlet,sarkar2008measurement,pileio2015real,pileio2017accessing,mamone2018nuclear,tourell2018singlet},
and detection of ligand--protein complexes that have weak binding~\cite{salvi2012boosting,buratto2014drug,buratto2014exploring}.
In applications for detecting ligand--protein complexes, the proton spins in the ligand can strongly interact with the nuclear spins in the protein because of the high gyromagnetic ratio of the proton.
To simulate the molecular motion of the systems and to obtain the correlation times more precisely, the molecular dynamics (MD) method was used.
Subsequently, the MD method was combined with QC calculations in nuclear spin relaxation time studies based on the Abragam--Redfield relaxation theory~\cite{peter2001calculation,case2002molecular,calero20151h,singer2017molecular,wildenberg2020computational}.
Furthermore, the relaxation time prediction of the nuclear spin-singlet state was performed by combining the MD simulation and QC calculations with a Kriging model~\cite{haakansson2017prediction} to calculate the relaxation time, mainly focusing on the spin-internal motion.

In this work, we predict the relaxation time of the singlet state composed of a proton spin pair and the longitudinal relaxation time for several small molecules in various solvents, while considering intermolecular dipolar interactions.
Next, we compare the expected and experimentally obtained relaxation times.
As our program calculates the contribution of each interaction, we can determine the most influential relaxation mechanism in nuclear spin relaxation.

The paper is organized as follows. 
In Sec. II, we briefly introduce analytical solutions for the relaxation rate derived from the Abragam--Redfield relaxation theory.
In Sec. III, we describe computational methods of our calculation.
In Sec. IV, we show our calculation results of the rotational correlation times and relaxation times of several molecules.
In Sec. V, we conclude our discussion.

\section{Theory}

\begin{table}[htbp]
    \centering
    \caption{Longitudinal relaxation rates $R_1$ and relaxation rates of the singlet state $R_{\mathrm{S}}$ for each relaxation mechanism.
    $\omega_i$ is the angular frequency of nucleus $i$.
    % $N_{\mathrm{in(ex)}}$ denotes the number of the nuclear spins in the target molecule (solvents) except for the spins $j$ and $k$.
    $\sigma^{i,\mathrm{a}}_{\alpha\beta}$ are the $\alpha\beta$ components of the antisymmetric part of the chemical shielding tensor in the molecular frame, which is the principal frame of the inertia tensor of the molecule.
    $\sigma^{i,\mathrm{s}}_{\alpha\beta}$ are the $\alpha\beta$ components of the symmetric part of the chemical shielding tensor in its principal frame.
    $C^i_{\alpha\beta}$ is the $\alpha\beta$ component of the spin--rotation tensor in the molecular frame.
    $||\cdot||$ denotes the Frobenius norm.
    $\mathrm{I}_{n}$ denotes the $n$ components of the molecular angular momentum.
    Here, we assume that nuclear spin $q$ and nuclear spins $j$ and $k$ are heteronuclear spins. In addition, we assume the first motion-limit regime for the spin--rotation (SR) interaction.
    }
    \renewcommand{\arraystretch}{1.3}
    \begin{tabular}{ccc} \hline \hline
        \hspace{5mm}$\Lambda$\hspace{5mm}   & $R_1^{\Lambda}$   & \hspace{30mm}$R_S^{\Lambda}$ \\ \hline
        DDjk    & \begin{tabular}{l} $\displaystyle \frac{1}{4}(c_{\mathrm{DD}}^{jk})^2\left(4J^{jk}_2(\Sigma\omega_{jk})+J^{jk}_1(\omega_j)\right)$ \\
        \end{tabular}
        & \renewcommand{\arraystretch}{1.3}
        0
        \vspace{1.5mm}\\
        % \vspace{1mm} 
        \cline{2-3} 
        in(ex)DD    & \begin{tabular}{l} 
        $\displaystyle\frac{1}{18}\sum_{q\in  N_{\mathrm{in(ex)}}} (c_{\mathrm{DD}}^{jq})^2I_q(I_q+1)$ \\
        \hspace{3mm}$\times \Bigl( 6J^{jq}_2(\Sigma\omega_{jq})+3J^{jq}_1(\omega_{j})+J^{jq}_0(\Delta\omega_{jq})$ \\
        \hspace{5mm}$+6J^{kq}_2(\Sigma\omega_{kq})+3J^{kq}_1(\omega_{k})+J^{kq}_0(\Delta\omega_{kq}) \Bigr)$ 
        \end{tabular}
        & \renewcommand{\arraystretch}{1.2}
        \begin{tabular}{l} $\displaystyle \frac{2}{27}\sum_{q\in N_{\mathrm{in(ex)}}}(c_{\mathrm{DD}}^{jq})^2I_q(I_q+1)\Bigl( 6J^{jq}_2(\Sigma\omega_{jq})$ \\
        $ +3J^{jq}_1(\omega_{j})+3J^{jq}_1(\omega_{q})+2J^{jq}_0(0)+J^{jq}_0(\Delta\omega_{jq})$ \\
        \hspace{1mm}$+6J^{kq}_2(\Sigma\omega_{kq})+3J^{kq}_1(\omega_{k})+3J^{kq}_1(\omega_{q})$ \\
        \hspace{1mm}$ +2J^{kq}_0(0)+J^{kq}_0(\Delta\omega_{jq}) - 6J^{jqkq}_2(\Sigma\omega_{jq}) $ \\ 
        \hspace{1mm} $-3J^{jqkq}_1(\omega_{j})-3J^{jqkq}_1(\omega_{q})-2J^{jqkq}_0(0) $ \\ 
        \hspace{1mm} $-J^{jqkq}_0(\Delta\omega_{jq}) -6J^{kqjq}_2(\Sigma\omega_{kq}) -3J^{kqjq}_1(\omega_{k}) $ \\ 
        \hspace{1mm} $-3J^{kqjq}_1(\omega_{q}) -2J^{kqjq}_0(0) -J^{kqjq}_0(\Delta\omega_{jq})\Bigr)$ 
        \vspace{1mm}
        \end{tabular}
        \\
        \cline{2-3}
        aCSA    & $\displaystyle \sum_{i=j,k}\frac{\gamma_i^2 B_0^2 }{3}\frac{\tau_1^{jk}}{1+(\tau_1^{jk}\omega_j)^2} ||\sigma^{i,\mathrm{a}}||^2$
        & \hspace{20mm}$\displaystyle \frac{2}{9}\gamma_j\gamma_k B_0^2\frac{\tau_1^{jk}}{1+(\tau_1^{jk}\omega_j)^2}||\Delta\sigma^{a}||^2$
        \vspace{5mm}\\
        sCSA    & $\displaystyle \sum_{i=j,k}\frac{\gamma_i^2 B_0^2}{20}\frac{\tau_2^{jk}}{1+(\tau_2^{jk}\Sigma\omega)^2}\left(\sigma_{zz}^{i,\mathrm{s}}\right)^2\left(3+\eta_i\right)^2$ 
        & \hspace{20mm}$\displaystyle \frac{2}{9}\gamma_j\gamma_k B_0^2\frac{\tau_2^{jk}}{1+(\tau_2^{jk}\Sigma\omega)^2}||\Delta\sigma^{s}||^2$\vspace{5mm}\\
        SR      & 
        $\displaystyle \frac{1}{9\hbar^2\tau_2^{jk}} \sum_{i=j,k} \sum_{n\in\{x,y,z\}}
        \left( ||C^{i,\mathrm{M}}||^2\right)\mathrm{I}_{n}^2$
        & \renewcommand{\arraystretch}{1} 
        \hspace{12mm} $\displaystyle \frac{1}{9\hbar^2 \tau_2^{jk}}\sum_{\alpha\in\{x,y,z\}}\sum_{\beta\in\{x,y,z\}}
        \left( ||\Delta C^{\mathrm{M}}_{\beta\alpha}||^2\right)
        \mathrm{I}_{\alpha}^2$ \vspace{5mm}\\
        \hline \hline
    \end{tabular}
    \leftline{$c_{\mathrm{DD}}^{ii'}=\frac{\hbar \mu_0 \gamma_i \gamma_{i'}}{4\pi}$; $\Sigma\omega_{ii'}=\omega_i+\omega_{i'}$; $\Delta\omega_{ii'}=\omega_i-\omega_{i'}$; $P_2(x)=\frac{3x^2-1}{2}$; $\eta_i=\frac{\sigma^{i,s}_{xx}-\sigma^{i,s}_{yy}}{\delta^i}$;}
    \leftline{$\Delta\sigma^{a(s)}=\sigma^{j,a(s)}-\sigma^{k,a(s)}$ ; $\Delta C^{\mathrm{M}}=C^{j,\mathrm{M}} - C^{k,\mathrm{M}}$}
  \label{tb:Relaxation_rates}
\end{table}

We introduce the relaxation rate of the longitudinal magnetization of two spin-1/2 nuclei $j$ and $k$ and that of the singlet state composed of the two nuclei derived from the Abragam--Redfield relaxation theory~\cite{abragam1961principles,redfield1965theory,kowalewski2006nuclear,pileio2020relaxation}.
In this study, we consider the following relaxation mechanisms: intra-pair dipolar~(DDjk), intramolecular dipolar~(inDD) except for DDjk, intermolecular dipolar~(exDD), anti-symmetric part of chemical shielding anisotropy~(aCSA), symmetric part of chemical shielding anisotropy~(sCSA), and spin--rotation~(SR).
\begin{eqnarray}
    \frac{1}{T_{N}} = R_{N}
    \simeq R_{\mathrm{DDjk}}+R_{\mathrm{inDD}}+R_{\mathrm{exDD}}+R_{\mathrm{CSA}}+R_{\mathrm{SR}}, \hspace{5mm} N=1,\mathrm{S}. \nonumber
\end{eqnarray}
Table~\ref{tb:Relaxation_rates} shows the analytical solutions of the relaxation rates used in this study.
Although most of these analytical solutions are shown in previous papers and books~\cite{abragam1961principles,redfield1965theory,kowalewski2006nuclear,pileio2011singletviaIDC,pileio2020relaxation}, the coefficient of the relaxation rate $R_{\mathrm{exDD}}$ and the relaxation rate $R_{\mathrm{SR}}$ obtained herein are slightly different from the values obtained previously.
The detailed derivations of these solutions are shown in Supplementary Material.
In this work, we neglect the relaxation rates due to other relaxation mechanisms, such as coherent leakage, and relaxation caused by paramagnetic impurities for the sake of simplicity.

$R_{\mathrm{DDjk}}$, the relaxation rate due to the dipolar interactions between spins $j$ and $k$ is expressed using the gyromagnetic ratios of the spins $\gamma_{j(k)}$;
reduced Planck constant, $\hbar$;
vacuum permeability, $\mu_0$;
and auto-correlation spectral density functions, $J^{jk}(\omega)$.
The auto-correlation spectral density functions of order $m$ are defined as
\begin{eqnarray}
    J^{jk}_m(\omega) &=& \int_0^{\infty} G^{jk}_m(\tau)\exp(-i\omega\tau) d\tau.
\end{eqnarray}
$G^{jk}_m(\tau)$ is the auto-correlation function of order $m$ with a time interval $\tau$.
The correlation function is defined as
\begin{eqnarray}
    G^{jk}_m(\tau) &=& 
    6\overline{r^{-3}_{jk}(0)D_{(0,m)}^{2}(\Omega^{jk}_{PL}(0))r^{-3}_{jk}(-\tau)D_{(0,m)}^{2,\ast}(\Omega^{jk}_{PL}(-\tau))}. \label{eq:autocorrelation}
\end{eqnarray}
Here, $D_{(0,m)}^{2}$ denotes the rank-2, component-$(0,m)$ of a Wigner matrix; $\Omega^{jk}_{PL}(t)(i=j,k)$ is the Euler angle set that connects the principal frame of the dipolar interaction and the laboratory frame at time $t$;
and $r_{jk}(\tau)$ is the distance between spins $j$ and $k$ at time $\tau$.

$R_{\mathrm{inDD}}$ and $R_{\mathrm{exDD}}$, the relaxation rates due to the intramolecular and intermolecular dipolar interactions except for DDjk are expressed using the quantities introduced in $R_{\mathrm{DDjk}}$;
spin quantum number of spin $q$, $I_q$;
and cross-correlation spectral density functions $J^{jqkq}(\omega)$.
Here, $q$ is the nuclear spin that interacts with spins $j$ and $k$.
The cross-correlation spectral density functions of order $m$ are defined as follows:
\begin{eqnarray}
    J^{jqkq}_m(\omega) &=& \int_0^{\infty} G^{jqkq}_m(\tau)\exp(-i\omega\tau) d\tau.
\end{eqnarray}
$G^{jqkq}_m(\tau)$ is the cross-correlation function of order $m$ with a time interval $\tau$.
The cross-correlation function is defined as:
\begin{eqnarray}
    G^{jqkq}_m(\tau) &=& 
    6\overline{r^{-3}_{jq}(0)D_{(0,m)}^{2}(\Omega^{jq}_{PL}(0))r^{-3}_{kq}(-\tau)D_{(0,m)}^{2,\ast}(\Omega^{kq}_{PL}(-\tau))}. \label{eq:crosscorrelation}
\end{eqnarray}
Next, the relaxation rates $R_{\mathrm{inDD}}$ $R_{\mathrm{exDD}}$ are calculated by summing the relaxation rates due to the dipolar interaction with all the spins in the target molecule ($q\in N_{\mathrm{in}}$) and solvent ($q\in N_{\mathrm{ex}}$).
$N_{\mathrm{in}}$ and $N_{\mathrm{ex}}$ are the numbers of spins in the target molecule and solvent, respectively.

$R_{\mathrm{CSA}}$, the relaxation rate due to the chemical shift anisotropy, is described using an external magnetic field $B_0$, the gyromagnetic ratios of the spin, the chemical shielding tensor $\sigma^{i}$ for spin $i$, and the first- and second-rank rotational correlation times $\tau_{1(2)}$.
$R_{\mathrm{SR}}$, the relaxation rate due to SR , is described using the second rotational correlational time; moment of inertia of the molecule $\mathrm{I}$;
and SR tensor of spin $i$, $C^{i}$.
In these relaxation mechanisms, the relaxation phenomenon is related to the molecular rotation and is associated with the first- and second-rank rotational correlation times for the target spins, $\tau_1^{jk}$ and $\tau_2^{jk}$, which are defined as
\begin{eqnarray}
    \tau_1^{jk} &=& \int_0^{\infty}\overline{\cos(\theta^{jk}(\tau))}d\tau, \label{eq:tau1}\\  
    \tau_2^{jk} &=& \frac{1}{2}\int_0^{\infty}\overline{3\cos(\theta^{jk}(\tau))^2-1}d\tau.\label{eq:tau2}
\end{eqnarray}
Here, $\theta^{jk}(\tau)$ denotes the angle between the vector connecting nuclear spin $j$ and nuclear spin $k$ at time $t=0$ and the vector at time $t=\tau$. The overline indicates ensemble averaging.

In summary, the relaxation rates can be calculated using the spectral density functions of the nuclear spins interacting with each other, first and second rotational correlation times, chemical shielding tensor, moment of inertia of the molecule, and SR tensor. 
It should be noted that, in this study, we neglect any cross-correlation between different mechanisms, such as DD--CSA , and coherent leakage mechanisms due to the chemical shift effects and scalar couplings for the sake of simplicity.

\section{Computational methods}

The quantities needed to calculate the relaxation rates, such as $\tau^{jk}_2$ and $J_m^{jk}(\omega)$, were obtained using the MD method and a QC calculation.
The computational details and the flowchart for the calculation are shown in Supplementary Material.

% MDではどんな計算をしたのか、QCではどんな計算をしたのかをIIで述べた各要素に関して述べる。
\subsection{Molecular dynamics method}

The spectral density functions and the rotational correlation times were computed via the MD method.
The MD method was performed using the AMBER 20 program package~\cite{gotz2012routine,salomon2013routine,le2013spfp}.
To calculate the spectral density functions in Eqs.~(\ref{eq:autocorrelation}) and~(\ref{eq:crosscorrelation}), the auto- and cross-correlation functions $G^{jqkq}(\tau)$ and $G^{jk}(\tau)$ had to be calculated.
The correlation functions were obtained from the vector between the two nuclear spins at time $t=0$ and the vector at $t=\tau$.
In the calculation of the correlation functions, except for $R_{\mathrm{exDD}}$, the vectors were extracted from the spins in the target molecule.
In the calculation for $R_{\mathrm{exDD}}$, the vectors were calculated using the spin in the target molecule and the spin in the solvent.
These vectors were extracted from the 200 NVE trajectories for 500~ps with different initial configurations, and the obtained correlation functions were averaged over the 200 NVE trajectories.
Subsequently, the spectral density functions were obtained by a Fourier transform of the correlation functions.
Finally, $R_{\mathrm{inDD}}$ and $R_{\mathrm{exDD}}$ were calculated using the sum of the spectral functions over all the spins in the target molecule $N_{\mathrm{in}}$ and solvents $N_{\mathrm{ex}}$.

The averaged correlation functions for the rotational correlation times $\overline{\cos(\theta^{jk}(\tau))}$ and $\overline{(3\cos(\theta^{jk}(\tau))^2-1)/2}$, which are shown in Eqs.~(\ref{eq:tau1}) and~(\ref{eq:tau2}), respectively, were also calculated from the 200 NVE trajectories.
The rotational correlation times were then obtained by fitting these correlation functions to $\exp(-t/\tau^{jk})$.

\subsection{Quantum Chemistry Calculation}
QC calculations were employed for the chemical shielding tensor, moment of inertia of the molecule, and SR tensor.
We used Gaussian 16 Revision A.03~\cite{ogliaro2016gaussian} and DALTON 2018.0 software~\cite{aidas2014d}.
The details of the calculation are shown in Supplementary Material.
The $R_{\mathrm{CSA}}$ and $R_{\mathrm{SR}}$ were calculated using the rotational correlation times and the aforementioned quantities.

\section{Results and discussion}

\begin{figure}[t]
    \centering
    \includegraphics[width=140mm]{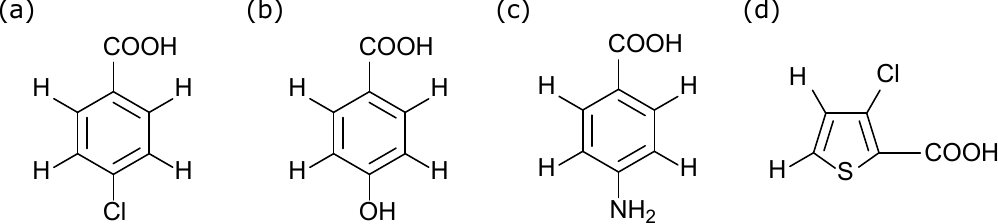}
    \caption{Molecular structures of (a) $p$-chlorobenzoic acid, (b) $p$-hydroxybenzoic acid, (c) $p$-aminobenzoic acid, and (d) 3-chlorothiophene-2-carboxylic acid.}
    \label{fig:molecular_structure}
\end{figure}

% ここでMDとQCの計算手法の詳細はsupplyに記載しましたとの旨を記載する。
Here, we present the calculation results of the rotational correlation times, correlation functions for the intermolecular dipolar interaction, and relaxation times obtained from the calculated relaxation rates.
In this work, we considered four molecules, including proton spin pairs with relaxation parameters measured in previous works: $p$-aminobenzoic acid~(PABA), $p$-chlorobenzoic acid~(PCBA), $p$-hydroxybenzoic acid~(PHBA), and 3-chlorothiophene-2-carboxylic acid (Cl-TC)~\cite{Pileio06,miyanishi2020long,kiryutin2019proton}.
These molecular structures are shown in Fig.~\ref{fig:molecular_structure}.
In PABA, PCBA, and PHBA, the relaxation of the two protons in the ortho and para positions was calculated.
In the case of Cl-TC, the relaxation of the two aromatic protons was calculated.
The proton at the carboxyl group was removed for the calculation in a system containing water or methanol.
The proton at the hydroxyl group was replaced by deuterium in a system containing water.
The protons attached to amide nitrogen were replaced by deuterium, and deuterium was added to amide for the calculation in a system containing water.
% \textcolor{blue}{In our MD simulations, the amide proton exchange is not taken into accounts.}

\begin{figure}
    \centering
    \includegraphics[width=160mm]{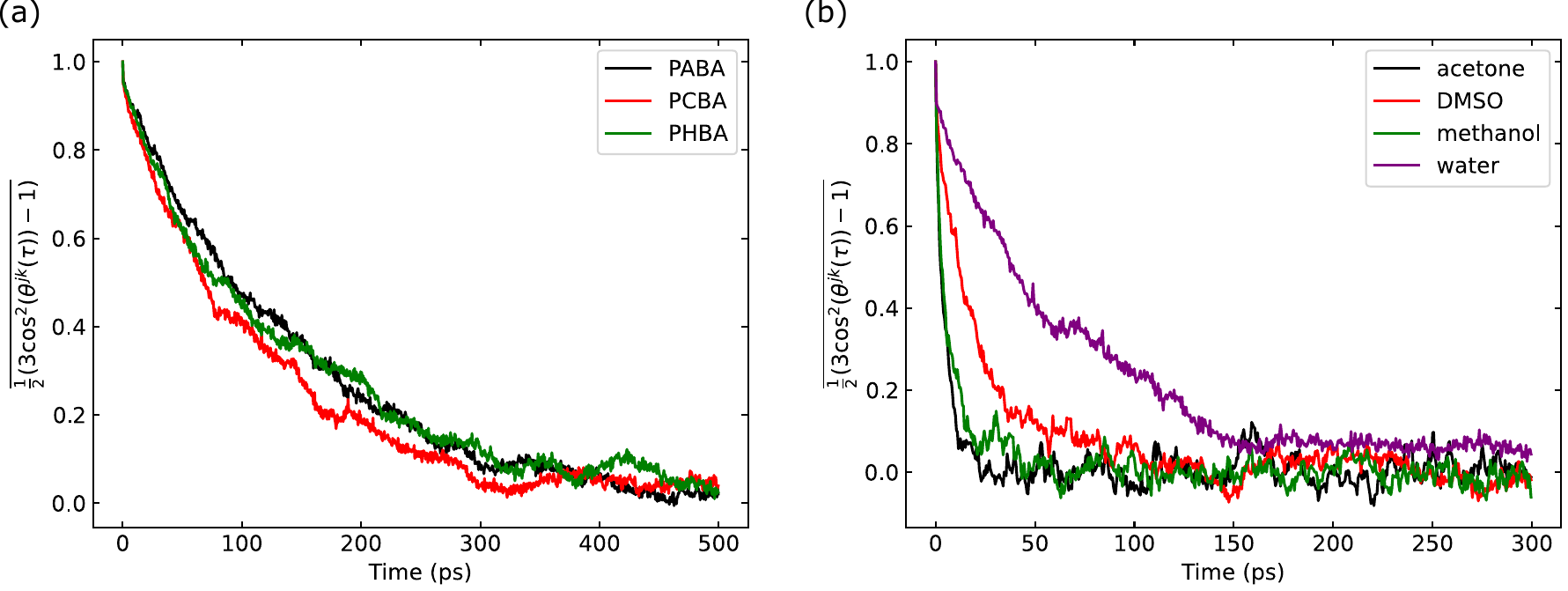}
    \caption{Simulation results of the second-rank rotational correlation function.
    (a) Simulation results for PABA~(black), PCBA~(green), PHBA~(red) in TIP4P water.
    (b) Simulation results for Cl-TC in acetone~(black), DMSO~(red), methanol~(green), and water~(purple).}
    \label{fig:rotation2}
\end{figure}

\subsection{Simulation results of the correlation functions and correlation times}
The simulation results of the averaged second-rank rotational correlation function $\overline{\frac{1}{2}(3\cos^2(\theta^{jk}(\tau))-1)}$ for the target protons are plotted in Fig.~\ref{fig:rotation2}.
The obtained correlation times are shown in Table~\ref{tb:correlation_times_rank1and2}.
The calculation results for PABA, PCBA, and PHBA are shown in Fig.~\ref{fig:rotation2}(a), and the calculated correlation times for these molecules are similar.
The calculation results for Cl-TC in several solvents are shown in Fig.~\ref{fig:rotation2}(b).
The ratios of the obtained second-rank rotational correlation times for these results, except for D$_2$O, are almost the same as the ratio of the kinetic viscosity of the solvent.
For Cl-TC in water, the rotational correlation time is significantly longer than the rotational correlation times in other solvents.
These results suggest that the coupling between ionized Cl-TC and water molecules increases the hydrodynamic radius of the target molecule. 
It is known that the rotational correlation time derived from the Stokes--Einstein--Debye relation is proportional to the cube of the hydrodynamic radius~\cite{bloembergen1948relaxation}.

\begin{figure}
    \centering
    \includegraphics[width=160mm]{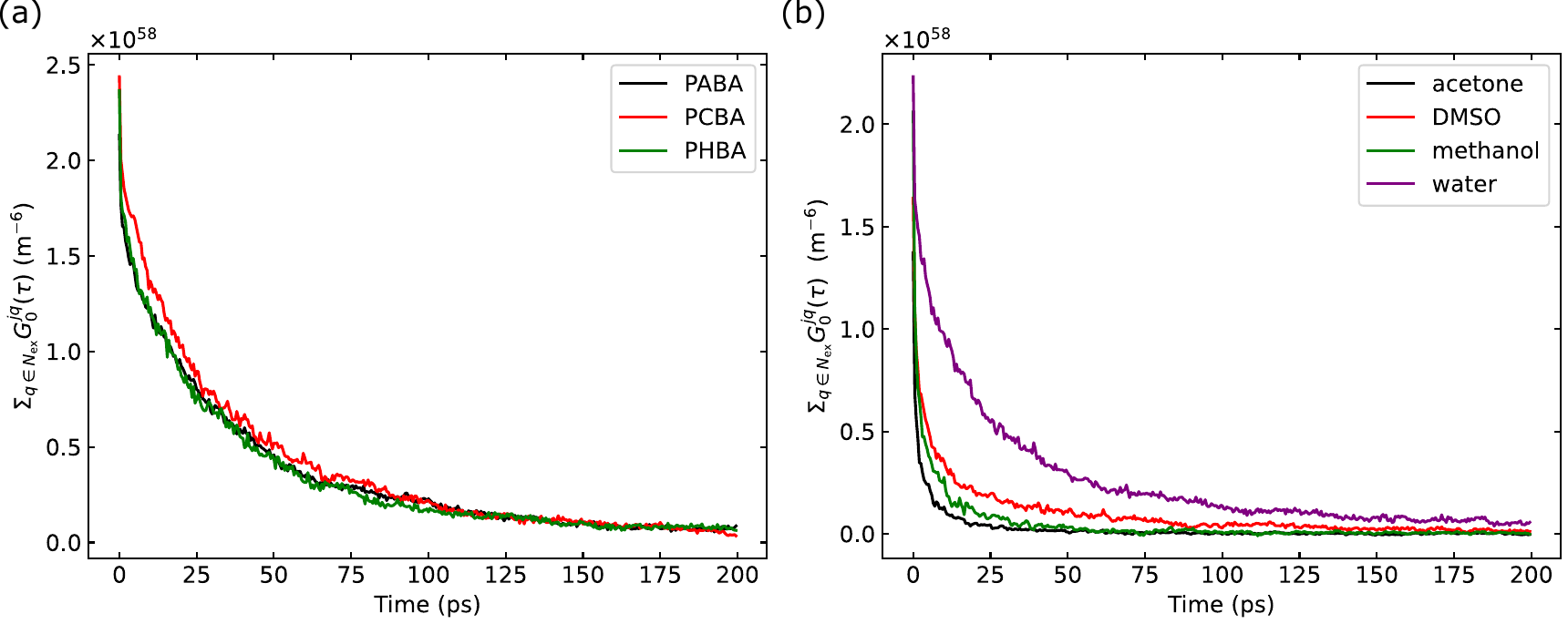}
    \caption{Simulation results of the auto-correlation function for the intermolecular dipolar interactions.
    (a) Simulation results for PABA~(black), PCBA~(red), PHBA~(green) in TIP4P water.
    (b) Simulation results for Cl-TC in acetone~(black), DMSO~(red), methanol~(green), and water~(purple).}
    \label{fig:interDD}
\end{figure}

The simulation results of the correlation function $\Sigma_{q\in N_{\mathrm{ex}}}G^{jq}_0(\tau)$ for the target proton pair are shown in Fig.~\ref{fig:interDD}.
The calculation results of the correlation functions for PABA, PCBA, and PHBA are shown in Fig.~\ref{fig:interDD}(a). As can be observed, the calculated correlation times are similar to those in the case of the second-rank rotational correlation time.
The calculation results of the correlation functions for Cl-TC in several solvents are shown in Fig.~\ref{fig:interDD}(b).
The correlation functions have different decay times depending on the solvent. The dependence of the decay time on the solvents is the same as that for the second-rank rotational correlation time of Cl-TC.

\begin{table}[t]
    \centering
    \caption{Calculated rank-1 and rank-2 correlation times.}
    \begin{tabular}{cccc} \hline \hline
        ~ molecule~                & ~~~solvent (viscosity~\cite{kiryutin2019proton})~~~&~~$\tau_1$~(ps)~&~~$\tau_2$~(ps)~ \\ \hline
    PABA~\cite{Pileio06}           &   water (0.89)                                    & 402            & 137 \\
    PCBA~\cite{miyanishi2020long}  &   water (0.89)                                    & 339            & 111 \\
    PHBA~\cite{Pileio06}           &   water (0.89)                                    & 415            & 136 \\
    Cl-TC~\cite{kiryutin2019proton}& acetone (0.39)                               & 12.2$\pm$0.2   & 4.6$\pm$0.1 \\     
    Cl-TC~\cite{kiryutin2019proton}& DMSO (1.8)                                   & 67.6$\pm$0.5   & 19.7$\pm$0.2\\
    Cl-TC~\cite{kiryutin2019proton}& methanol (0.68)                              & 21.5$\pm$0.3   & 6.7$\pm$0.1\\
    Cl-TC~\cite{kiryutin2019proton}& water (0.89)                                      & 185            & 63.9$\pm$0.4\\
    \hline \hline
    \end{tabular}
  \label{tb:correlation_times_rank1and2}
\end{table}

\subsection{Calculation results of the relaxation time}

The calculated relaxation times for the longitudinal magnetization and the singlet state are compared in Fig.~\ref{fig:R1} and Fig.~\ref{fig:RS}.
The detailed values are provided in Supplementary Material.
Regarding the error in our calculations, the statistical errors in the MD calculations were estimated using the jackknife method for dipolar interactions~\cite{efron1981jackknife}.
The statistical errors for the relaxation rate due to the CSA and SR interactions were estimated using the error in the fitting for the rotational correlation times.
The relaxation rate due to the SR interaction is not shown in Fig.~\ref{fig:R1} and Fig.~\ref{fig:RS} because the contribution of SR is considerably smaller than that of the other mechanisms.

\begin{figure}
    \centering
    \includegraphics[width=160mm]{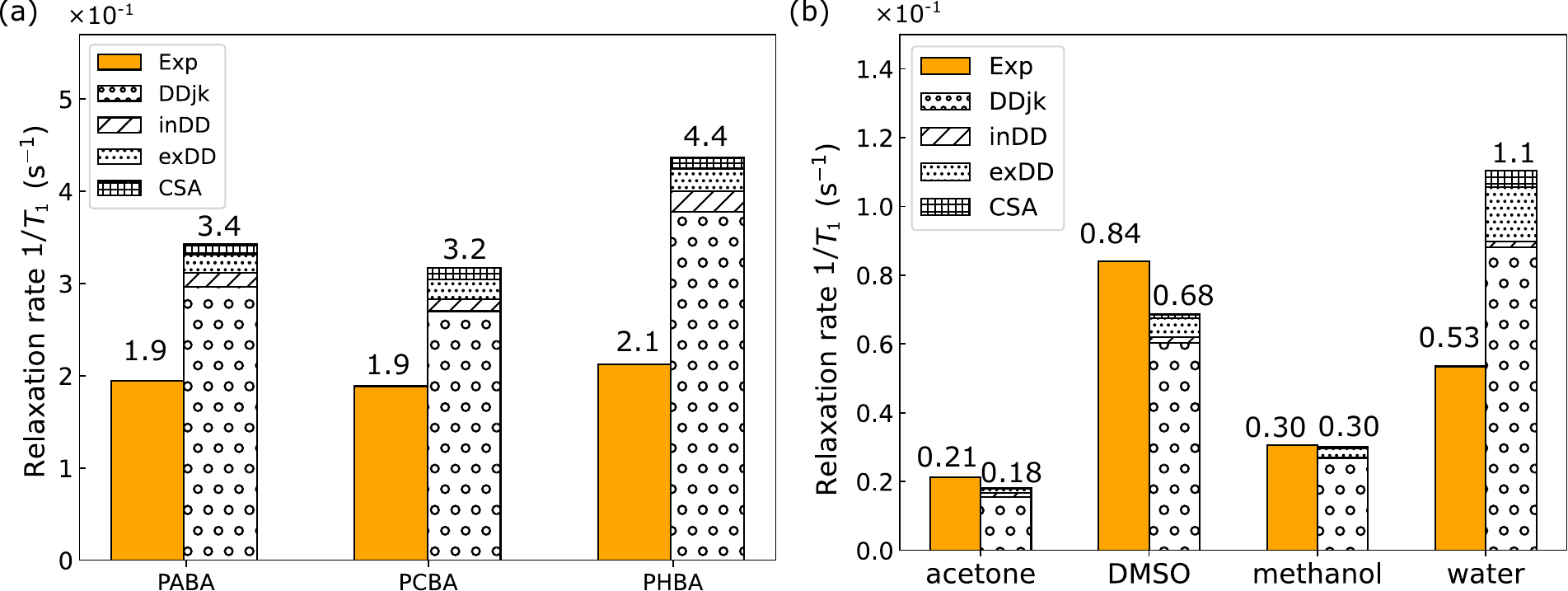}
    \caption{Experimental and calculation results of the relaxation time for longitudinal magnetization.
    (a) Results for PABA, PCBA, and PHBA in TIP4P water.
    (b) Results for Cl-TC in acetone, DMSO, methanol, and water.}
    \label{fig:R1}
\end{figure}

\begin{figure}
    \centering
    \includegraphics[width=160mm]{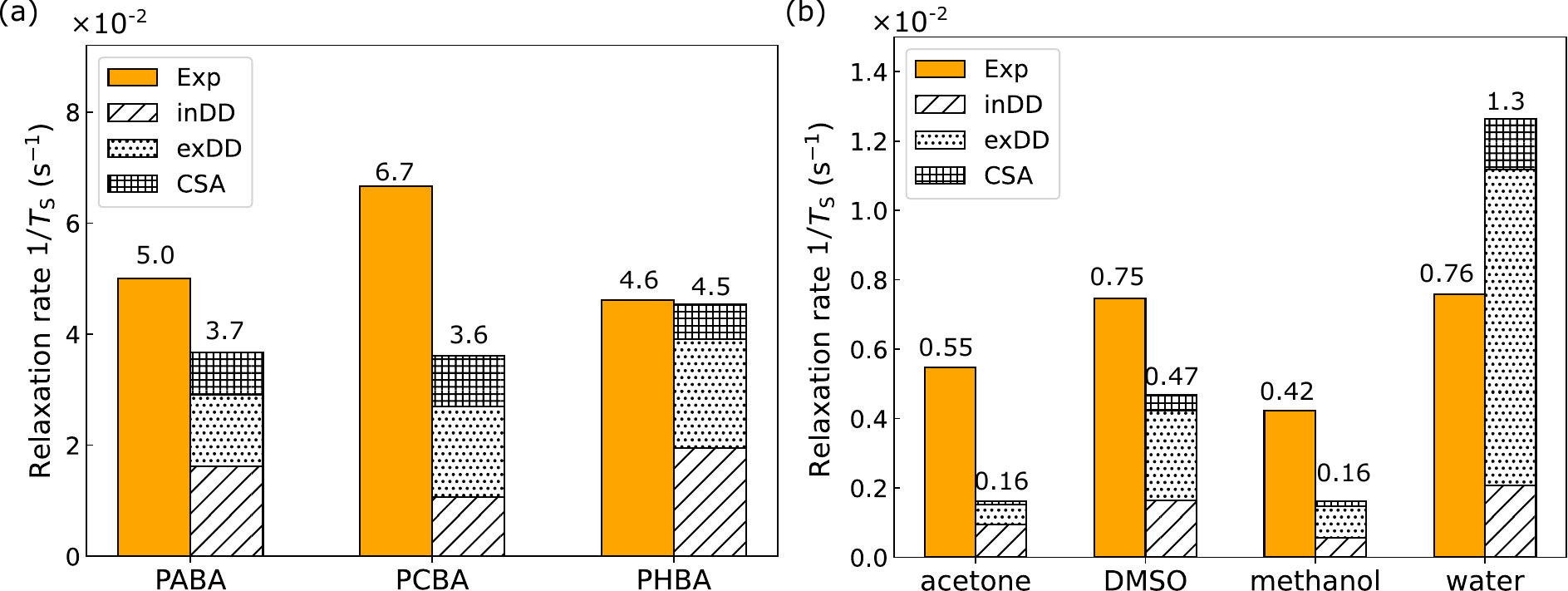}
    \caption{Experimental and calculation results of the relaxation time for the long-lived state.
    (a) Results for PABA, PCBA, and PHBA in TIP4P water.
    (b) Results for Cl-TC in acetone, DMSO, methanol, and water.}
    \label{fig:RS}
\end{figure}

Overall, the relaxation time can be estimated with some accuracy for longitudinal magnetization.
The error between the experimental and calculated results is smaller than 52~\%.
In contrast, the estimated value of the relaxation time for the singlet state and the calculated values are of the same order.
We discuss these results for each calculation below.

(i) PABA, PCBA, and PHBA:
The dominant longitudinal relaxation mechanism is provided by the intra-pair dipolar interaction DDjk, the contribution of which is larger than 85~\%.
The contributions from the inDD, exDD, CSA, and SR terms are significantly smaller by one order of magnitude than the DDjk term.
In comparison with the simulated results, for PABA, the experimental $T_1$ = 5.15~s is underestimated by 49~\% at 2.43~s;
for PCBA, the experimental $T_1$ = 5.3~s is underestimated by 43~\% at 2.92~s;  
and for PHBA, the experimental $T_1$ = 4.71~s is underestimated by 42~\% at 3.09~s.

For the relaxation time of the singlet state, the relaxation rates due to the inDD, exDD, and CSA interactions are of the same order.
These results show that exDD contributes significantly to the relaxation of the singlet state composed of proton spins.
In comparison with the simulated results, for PABA, the experimental $T_{\mathrm{S}}$ = 20~s is overestimated by 36~\% at 27.2~s;
for PCBA, the experimental $T_{\mathrm{S}}$ = 15~s is overestimated by 85~\% at 27.7~s; and
for PHBA, the experimental $T_{\mathrm{S}}$ = 21.7~s is overestimated by 1.6~\% at 22.1~s.
It should be noted that the relaxation time for PCBA was measured in a dissolution DNP experiment, in which a hyperpolarized sample was dissolved into a hot solvent and the sample contained dissolved oxygen.
Dissolved oxygen is a paramagnetic substance, leading to nuclear spin relaxation~\cite{erriah2019experimental}.
The simulation of the relaxation due to dissolved oxygen would make the prediction of the relaxation times more accurate.
In our MD simulations, the amide proton exchange was not taken into account.
For PABA, the simulation of the relaxation due to this exchange process could improve the accuracy of the relaxation time prediction.

(ii) Cl-TC in acetone, DMSO, methanol, and water. 
The dominant longitudinal relaxation mechanism is provided by the intra-pair dipolar interaction DDjk, and the obtained longitudinal relaxation times $T_1$ are consistent with the experimental results.
The contributions from the inDD, exDD, CSA, and SR terms are smaller by one order  of magnitude than that from the DDjk term as in the case of benzoic acid derivatives.
In comparison with the simulated results, for acetone, the experimental $T_1$ = 47~s is overestimated by 17~\% at 55~s; for DMSO, the experimental $T_1$ = 11.9~s is overestimated by 23~\% at 14.6~s; for methanol, 
the experimental $T_1$ = 32.8~s is overestimated by 1.5~\% at 33.3~s;
and for water, the experimental $T_1$ = 18.7~s is overestimated by 52~\% at 9.05~s.

For the relaxation time of the singlet state, the relaxation rates due to the inDD, exDD, and CSA interactions are of the same order.
The inDD singlet relaxation results from proton or deuterium at the carboxyl group and chloride.
In the calculation result for water, the dominant relaxation mechanism is provided by exDD.
It is presumed that ionization in water strengthens the coupling between Cl-TC and water.
Regarding the calculation result of the exDD interaction in other solvents, the relaxation rate is proportional to the viscosity of the solvents.
These calculation results show that exDD significantly contributes to the relaxation of the singlet state composed of protons in these solvents.
In comparison with the simulated results, for acetone, the experimental $T_{\mathrm{S}}$ = 183~s is overestimated by 237~\% at 616~s;
for DMSO, the experimental $T_{\mathrm{S}}$ = 134~s is overestimated by 59~\% at 214~s;
for methanol, the experimental $T_{\mathrm{S}}$ = 236~s is overestimated by 160~\% at 615~s; and
for water, the experimental $T_{\mathrm{S}}$ = 132~s is underestimated by 40~\% at 79.1~s.

In both calculations (i) and (ii), the exDD relaxation has several contributions to the longitudinal relaxation and large contributions from 35\% to 72\% to the relaxation of the singlet state.
Considering the gyromagnetic ratio and spin quantum number of hydrogen and deuterium, the value of $R_{\mathrm{exDD}}$ for H$_2$O is approximately 16 times larger than that for D$_2$O.
These results may explain the difference between the relaxation times of Cl-TC in D$_2$O and H$_2$O~\cite{kiryutin2019proton}.
This exDD relaxation mechanism might be a reason for the high contrast between the lifetimes of $T_{\mathrm{S}}$ of the bound and free forms~\cite{salvi2012boosting,buratto2014drug,buratto2014exploring,miyanishi2020long}.
For a more precise estimation of $R_{\mathrm{exDD}}$, the relaxation mechanism due to proton exchange in the carboxyl, hydroxyl, and amide groups can be considered using a reactive force field~\cite{ufimtsev2007charged,senftle2016reaxff} or QM/MM, but they are beyond the scope of this study.

% 計算結果が含むであろう誤差要因
The NMR relaxation properties depend on the model of the solvents and deuteration of the solvents~\cite{haakansson2017prediction}.
In our MD simulations, we used TIP4P~\cite{jorgensen1983comparison} for water.
Thus, using other water models (SPC/HW~\cite{spchw2001}, TIP3P~\cite{TIP3P}, TIP4P-Ew~\cite{TIP4P-Ew}, TIP4P2005f~\cite{TIP4P/2005f}) and/or the deuteration of the solvents may improve the accuracy of the relaxation rate estimations. 
By performing calculations of various combinations of molecules and water models, we may empirically suggest an optimal water model for a group of molecules.

The CSA interaction has some contribution, and higher accuracy of the QC calculation would improve the estimated relaxation rate in the calculation of singlet relaxation.
The QC calculation using the system containing the target molecule and the solvent molecules within several angstroms of the target molecule would yield more accurate results than the calculation using the polarizable continuum model.
In addition, QM/MM simulations would improve the accuracy of the QC calculations.

\section{Conclusion}
We predicted the relaxation time for the longitudinal magnetization and the long-lived state using the MD method and a QC calculation.
In these calculations, we considered the relaxation mechanisms of the intra-pair dipolar, intramolecular dipolar, intermolecular dipolar, chemical shift anisotropy, and SR interaction.
For example, we calculated the relaxation times for aromatic protons of PABA, PCBA, and PHBA and the aromatic protons of Cl-TC in acetone, DMSO, methanol, and water.
Using our calculation program, we estimated the longitudinal relaxation time with an error within 52\% and found that the intra-pair dipolar interaction accounted for more than 80\% of the relaxation rate for those systems.
We analyzed the calculation results for each relaxation mechanism to determine the components that contribute more to nuclear spin relaxation. 
In our simulation, it turned out that the intermolecular dipolar interaction between the proton spin and deuterium in solvents has a considerable contribution to the relaxation of a singlet state composed of proton spin pairs.

Our relaxation time prediction method will help in the future search for new molecular structures with long relaxation times of the singlet state for applications using long-lived states, such as long-term preservation of enhanced polarization~\cite{vasos2009long,ahuja2010proton,kiryutin2012creating,eills2021singlet},
molecular tag to probe slow dynamic processes~\cite{sarkar2007singlet,sarkar2008measurement,pileio2015real,pileio2017accessing,mamone2018nuclear,tourell2018singlet},
and detection of ligand--protein complexes that have weak binding~\cite{salvi2012boosting,buratto2014drug,buratto2014exploring}.
In particular, our relaxation time prediction method will facilitate the search for new molecular structures with long longitudinal relaxation times $T_1$~\cite{Korenchan2017DicarboxylicAcidsPHsensor,nonaka2013platform,kondo2020design}.
We believe that this method will open up new applications in NMR/MRI spectroscopy.

\begin{acknowledgments}
This work was supported by the MEXT Quantum Leap Flagship Program (MEXT Q-LEAP), CREST (JST Grant No. JPMJCR1672), and PREST (grant No. JPMJPR18G5).
KM has been supported by JSPS KAKENHI No. 19J10976 and the Program for Leading Graduate Schools: Interactive Materials Science Cadet Program.

\end{acknowledgments}

\newpage

\section*{Supplementary Material}

\subsection*{Detailed calculation for nuclear spin relaxation}

We derive the relaxation rate of the longitudinal magnetization and singlet state composed of two homonuclear spin-1/2, $j$ and $k$, based on Abragam–-Redfield relaxation theory~\cite{abragam1961principles,redfield1965theory,kowalewski2006nuclear,pileio2020relaxation}.
We consider four relaxation mechanisms, intramolecular dipolar interaction, intermolecular dipolar interaction, chemical shift anisotropy, and spin--rotation.
The Hamiltonian of the spin system is written as below:
\begin{eqnarray}
    H(t)
    &=& H_0 + H_1(t) \nonumber \\
    &=& H_0 + H\u{DD}(t) + H\u{CSA}(t) + H\u{SR}(t).
\end{eqnarray}
$H_0$ is the time-independent part of the spin Hamiltonian and mainly written by the Zeeman interaction and the spin-spin coupling.
$H\u{DD}, H\u{CSA}, H\u{SR}$ is the Hamiltonian describing the dipolar~(DD), chemical shift anisotropy~(CSA), and spin--rotation~(SR) interaction.
It is convenient to introduce the expression using spherical harmonics when calculating the nuclear spin relaxation due to the time-dependent Hamiltonians~\cite{abragam1961principles,kowalewski2006nuclear,pileio2009theory,pileio2020relaxation}.
Each of the interaction Hamiltonians ($\Lambda$) can be written using an interaction strength constant $c^{\Lambda}$, irreducible spherical tensor operators $T^{\Lambda}_{(l,m)}$, and spherical harmonics $A^{\Lambda}_{(l,m)}(t)$ as
\begin{eqnarray}
    H\u{\Lambda}(t) &=& c^{\Lambda}\sum^{l}_{m=-l} (-1)^m A^{\Lambda}_{(l,m)}(t) T^{\Lambda}_{(l,-m)},
    \label{eq:Spherical_representation_Hamiltonian}
\end{eqnarray}
where the rank $l$ depends on the interaction Hamiltonian.
The values for each of the interactions are summarized in Table~\ref{tb:Spherical_representation_Hamiltonian}.
In the following calculations, the Hamiltonians are represented in three coordinate systems, laboratory frame~(L), molecular frame~(M) and principal frame~(P).
The laboratory frame is a frame whoze z-axis is parallel to the external magnetic field.
The molecular frame is a frame that diagonalizes the inertia tensor of a molecule, and this frame is useful for discussing an interaction between the nuclear spin angular momentum and molecular rotation.
The principal frame is the reference frame defined for a rank-2 tensor, and in this frame the off-diagonal components of the interaction tensor becomes zero.
In the following, the frame of the spin operator part is fixed to the laboratory frame.

\begin{table}
    \centering
    \caption{An interaction strength constant $c^{\Lambda}$, irreducible spherical tensor operators $T^{\Lambda}_{(l,m)}$, and spherical harmonics $A^{\Lambda}_{(l,m)}(t)$ for each interaction mechanism.
    DD$ii'$ is the dipolar interaction between spin $i$ and $i'$, 
    sCSA is the interaction due to symmetric part of the chemical shielding anisotropy,
    aCSA is the interaction due to antisymmetric part of the chemical shielding anisotropy,
    and SR is the spin--rotation interaction.
    $\mu_0$ is the vacuum permeability. 
    $\hbar$ is the reduced Planck constant.
    $\gamma_i$ is the gyromagnetic ratio of the nuclear spin $i$.
    $r_{ii'}$ is the distance between the two nuclei $i$ and $i'$.
    $I_{\alpha i}, (\alpha\in\{x,y,z\})$ is $\alpha$ component of the spin vector of the $i$ spin.
    $I_{\pm i}=(I_{xi}\pm I_{yi})/2$ is the raisng and lowering operator.
    $\sigma_{\alpha\beta}^{i,\mathrm{s(a)}}, (\alpha, \beta \in \{x,y,z\})$ is the Cartesian components of the symmetric (antisymmetric) chemical shift tensor for spin $i$ in the principal (molecular) frame.
    $C_{\alpha\beta}^i$ denotes the Cartesian components of the spin--rotation tensor for spin $i$ in the molecular frame.}
    \begin{tabular}{ccccccccc} \hline \hline
        $\Lambda$    & $c^{\Lambda}$ & $l$ & $m$ & $A_{(l,m)}^{\Lambda,\mathrm{P}}(t)$ &  $T^{\Lambda}_{(l,m)}$   \\ \hline
        DD$ii'$&$\frac{\hbar\mu_0\gamma_i\gamma_{i'}}{4\pi}$ &2&0     & $\sqrt{6}r^{-3}_{ii'}(t)$ & $\frac{1}{\sqrt{6}}(3I_{zi}I_{zi'}-\mathbf{I}_i\cdot\mathbf{I}_{i'})$ \\ 
              &                                             & &$\pm1$& 0                  &$\mp\frac{1}{2}(I_{zi}I_{\pm i'}+I_{\pm i}I_{z i'})$  \\
              &                                             & &$\pm2$& 0                              &$\frac{1}{2}I_{\pm i}I_{\pm i'}$  \\
        sCSA   &$\gamma_i$                                   &2&   0  &$\sqrt{\frac{3}{2}}\delta^{i}$ &$\sqrt{\frac{2}{3}}B_0I_{zi}$ \\
              &                                             & &$\pm1$&       0                                       &$\mp\frac{1}{2}B_0 I_{\pm i}$ \\
              &                                             & &$\pm2$&$\frac{\eta^i}{2}\delta^{i}$   &   0  \\
        aCSA   &$\gamma_i$                                   &1&   0  & $\sqrt{2}\sigma^{i,\mathrm{a}}_{xy}$   &  0 \\
              &                                             & &$\pm1$&$\sigma_{zx}^{i,\mathrm{a}}\pm i\sigma_{zy}^{i,\mathrm{a}}$ &  $\pm\frac{1}{2}B_0 I_{\pm i}$  \\
        SR     &1                                            &1&   0  &  $\sum_{\alpha}C_{z\alpha}^{i}J_{\alpha}(t)$   &  $I_{zi}$   \\
              &                                             & &$\pm1$& $-\frac{1}{\sqrt{2}}\sum_{\alpha}(C^{i}_{x\alpha}J_{\alpha}(t)\pm iC^{i}_{y\alpha}J_{\alpha}(t))$ &$\pm \frac{1}{\sqrt{2}}I_{\pm i}$  \\
        \hline \hline
    \end{tabular}
    \leftline{$\eta^{i}=(\sigma^{i,\mathrm{s}}_{yy}-\sigma^{i,\mathrm{s}}_{xx})/\sigma^{i,\mathrm{s}}_{zz}$, $\delta^i=\sigma_{zz}^{i,\mathrm{s}}-(\sigma_{xx}^{i,\mathrm{s}}+\sigma_{yy}^{i,\mathrm{s}})/2$} 
    \label{tb:Spherical_representation_Hamiltonian}
\end{table}

The master equation of relaxation of nuclear spins can be derived from Liouville--von Neumann equation using perturbative approach~\cite{abragam1961principles,redfield1965theory,kowalewski2006nuclear,pileio2020relaxation} and expressed as
\begin{eqnarray}
    \frac{d}{dt}\overline{\tilde{\rho}(t)}
    &=& -\int_0^{\infty}\left[\overline{\tilde{H}_1(0),[\tilde{H}_1(-\tau)}, \overline{\tilde{\rho}(t)}]\right]d\tau \nonumber \\
    &=& -\Gamma\overline{\tilde{\rho}(t)} \nonumber \\
    &=& -(\Gamma\u{DD}+\Gamma\u{CSA}+\Gamma\u{SR})\overline{\tilde{\rho}(t)}
    \label{eq:relax_master_equation}
\end{eqnarray}
where $\rho(t)$ is the density matrix operator for the nuclear spins, the overbar represents an average over the molecular ensemble, the tilde indicates that the operator is written in the interaction frame of the time-independent Hamiltonian and $\Gamma_{\Lambda}$ indicates the relaxation superoperator for the interaction $\Lambda$.

In the followings, we consider the master equation of relaxation for each relaxation mechanisms.
It should be noted that we neglect any cross-correlation between different mechanisms such as DD–CSA, coherent leakage mechanisms due to the chemical shift effects, and scalar-coupling for the sake of simplicity.

\subsubsection{Dipolar interaction}
All possible relaxation superoperators due to the dipolar interactions are written as
\begin{eqnarray}
    \Gamma_{\mathrm{DD}}=&
    \Gamma^{jk,jk}_{\mathrm{DDjk}}+
    \sum_{q\in N_{\mathrm{in}}}\sum_{rs}\sum_{tu}\Gamma^{rs,tu}_{\mathrm{inDD}}+
    \sum_{q\in N_{\mathrm{ex}}}\sum_{rs}\sum_{tu}\Gamma^{rs,tu}_{\mathrm{exDD}}, \nonumber \\
    &rs,tu\in\{jq, kq\}.
\end{eqnarray}
Here, $\Gamma^{rs,tu}_{\Lambda}$ is the relaxation superoperators for the dipolar interaction between spin pairs $rs$ and $tu$,
$N_{\mathrm{in}}$ denotes all nuclear spins except spins $j$ and $k$ in the target molecule, and $N_{\mathrm{ex}}$ denotes the all nuclear spins of the solvent molecules.

Using Eq.~(\ref{eq:relax_master_equation}) and the values in Table~\ref{tb:Spherical_representation_Hamiltonian}, the relaxation superoperator of the dipolar interaction DD$_\Lambda$ is written as
\begin{eqnarray}
    \Gamma^{rs,tu}_{\mathrm{DD_\Lambda}} 
    &=&
    -c^{rs,\mathrm{DD_\Lambda}}c^{tu,\mathrm{DD_\Lambda}}\sum^{2}_{m,m^{\prime}=-2}(-1)^{(m+m^{\prime})} \nonumber \\
    &\times& \int_0^{\infty}6
    \overline{
        r_{rs}^{-3}(0)D_{(0,m)}^{2}(\Omega^{rs}_{\mathrm{PL}}(0)) 
        \tilde{T}^{rs,\mathrm{DD_\Lambda}}_{(2,-m)}
        r_{tu}^{-3}(-\tau)D_{(0,m')}^{2,\ast}(\Omega^{tu}_{\mathrm{PL}}(-\tau)) 
        \tilde{T}^{tu,\mathrm{DD_\Lambda},\dagger}_{(2,-m^{\prime})}} 
    d\tau, \nonumber \\
\end{eqnarray}
where $D_{(0,m)}^{2}(\Omega_{\mathrm{PL}}(t))$ is the Wigner matrices to change the reference frame of the spatial part.
$\Omega_{\mathrm{PL}}(t)=(\alpha_{\mathrm{PL}}(t), \beta_{\mathrm{PL}}(t), \gamma_{\mathrm{PL}}(t))$ is the Euler angles between the laboratory frame~(L) and the principal frame of the Hamiltonian~(P).

The Hermite conjugate of the spherical tensor commutation superoperator in the rotating frame can be evaluated as $\tilde{T}^{\dagger}_{(l,-m)}=(-1)^{m}\tilde{T}_{(l,m)}$, and the spherical tensor operators in the interaction frame are written as~\cite{pileio2011singlet}
\begin{eqnarray}
    \tilde{T}^{ii'}_{(2,m)} &=\sum_{\mu=1}^{3-|m|}\exp[i\omega_m^{ii',\mu} t]S^{ii',\mu}_{m},
    \label{eq:DDintra_interaction_frame}
\end{eqnarray}
where $\omega_m^{ii',\mu}$ and $S_m^{ii',\mu}$ are shown in Table~\ref{tb:DDintra_interaction_frame}, and when the spins $p$ and $q$ are homonuclear spins, $\omega_m^{\mu}=m\omega_p$ holds.
\begin{table}[t]
    \centering
    \caption{Symbols in Eq.~(\protect\ref{eq:DDintra_interaction_frame})}
    \begin{tabular}{cccccccc} \hline \hline
            & \multicolumn{3}{c}{$S^{ii',\mu}_{m}$} & & \multicolumn{3}{c}{$\omega^{ii',\mu}_{m}$}  \\ \cline{2-4} \cline{6-8} 
        $m$ & $\mu=1$ & $\mu=2$ & $\mu=3$           & & $\mu=1$ & $\mu=2$ & $\mu=3$     \\ \hline
        -2  &$\frac{1}{2}I_{-i}I_{-i'}$ &- &-    & &$-\omega_i -\omega_{i'}$ &- &-   \\
        -1  &$\frac{1}{2}I_{{zi}} I_{-i'}$& $\frac{1}{2}I_{-i} I_{zi'}$ &- & & $-\omega_{i'}$ & $-\omega_{i}$ &  - \\
        0            &$\frac{1}{2\sqrt{6}}(4I_{zi}I_{zi'})$   &$-\frac{1}{2\sqrt{6}}(I_{+i}I_{-i'})$  &$-\frac{1}{2\sqrt{6}}(I_{-i}I_{+i'})$  & & 0 &$\omega_i -\omega_{i'}$ &$\omega_{i'} -\omega_{i}$   \\
        1      &$-\frac{1}{2}(I_{zi} I_{+i'})$    &$-\frac{1}{2}(I_{+i} I_{zi'})$ &-  & & $\omega_{i'}$ & $\omega_i$   \\
        2   &$\frac{1}{2}I_{+i}I_{+i'}$ &- &-    & &$\omega_i +\omega_{i'}$ &- &- \\
        \hline \hline
    \end{tabular}
    \renewcommand{\arraystretch}{1}
  \label{tb:DDintra_interaction_frame}
\end{table}
From this relation, the relaxation superoperator becomes
\begin{eqnarray}
    \Gamma^{rs,tu}_{\mathrm{DD_\Lambda}} 
    &=& -c^{rs,\mathrm{DD_\Lambda}}c^{tu,\mathrm{DD_\Lambda}}\sum^{2}_{m=-2}\sum_{\mu=1}^{3-|m|}(-1)^mJ(\omega)
    T^{rs}_{(2,-m)} S^{tu,\mu}_{m}.
\end{eqnarray}
Here, $J^{rs,tu}_{m}(\omega)$ is the spectral fanction defined as
\begin{eqnarray}
    J^{rs,tu}_{m}(\omega)
    &=& 
    \int_0^{\infty}
    G^{rs,tu}_{mm}(\tau)
    \exp[i\omega\tau]d\tau,
\end{eqnarray}
where $G^{rs,tu}_{mm}(\tau)=6\overline{r_{rs}^{-3}(0)D_{(0,m)}^{2}(\Omega^{rs}_{\mathrm{PL}}(0))r_{tu}^{-3}(-\tau)D_{(0,m)}^{2,\ast}(\Omega^{tu}_{\mathrm{PL}}(-\tau))}$ is the correlation function. 
Here, we describe the auto-correlation spectral function and the auto-correlation function as $J^{rs}_{m}(\omega)$, $G^{rs}_{mm}(\tau)$.
Then, the longitudinal relaxation rate $R_1$ due to the dipolar interactions become
\begin{eqnarray}
    R^{\mathrm{DD_\Lambda}}_{1}&=- 
     \frac{1}{2} \mathrm{Tr}\left[(I_{zj}+I_{zk})
     \Gamma^{rs,tu}_{\mathrm{DD_\Lambda}}
     (I_{zj}+I_{zk})\right] \nonumber \\
    &=
    \begin{cases}
    \frac{1}{4}(c^{jk,\mathrm{DD_\Lambda}})^2(4J_2(\Sigma\omega)+J_1(\omega_j)) \\
    \hspace{20mm}\text{($rs=tu$, and spins $r$, $s$, and $j$ are homonuclear spins.)} \\
    \frac{1}{18}I_q(I_q+1)(c^{rs,\mathrm{DD_\Lambda}})^2
    (6J_2^{rs}(\Sigma\omega_{rs})+3J_1^{rs}(\omega_r)+J_0^{rs}(\Delta\omega_{rs})) \\ \hspace{20mm}\text{($rs=tu=jq$ or $kq$, where spins $j$ and $k$ and $q$ are heteronuclear spin.)} \\
    0 \hspace{18mm}\text{(otherwise)}
    \end{cases}
    \label{eq:R1_inDD}
\end{eqnarray}
The relaxation rate of the nuclear singlet state $R_{\mathrm{S}}$ due to the dipolar interactions become
\begin{align}
    R^{\mathrm{DD_\Lambda}}_{\mathrm{S}}&=- 
     \frac{4}{3} \mathrm{Tr}\left[(\mathbf{I}_j\cdot\mathbf{I}_k)
     \Gamma^{rs,tu}_{\mathrm{DD_\Lambda}}
     (\mathbf{I}_j\cdot\mathbf{I}_k)\right] \nonumber\\
    &=
    \begin{cases}
    \frac{2}{27}I_q(I_q+1)(c^{rs,\mathrm{DD_\Lambda}})^2
    (6J_2^{rs}(\Sigma\omega_{rs})+3J_1^{rs}(\omega_r)+3J_1^{rs}(\omega_s)+2J_0^{rs}(0)+J_0^{rs}(\Delta\omega_{rs}))\\
    \hspace{40mm}\text{($rs=tu=jq$ or $kq$)}\\
    -\frac{2}{27}I_q(I_q+1)c^{rs,\mathrm{DD_\Lambda}}c^{tu,\mathrm{DD_\Lambda}}\\
    \hspace{20mm}(6J_2^{rs,tu}(\Sigma\omega_{rs})+3J_1^{rs,tu}(\omega_r)+3J_1^{rs,tu}(\omega_s)+2J_0^{rs,tu}(0)+J_0^{rs,tu}(\Delta\omega_{rs}))\\
    \hspace{40mm}\text{($(rs,tu)=(jq,kq)$ or $(kq,jq)$)}. \\
    0 \hspace{40mm}\text{(otherwise)}\\
    \end{cases}.
\end{align}
Here, the spin $q$ is the nuclear spin either in the target molecule or the solvent.

\subsubsection{Chemical shift anisotropy}
Next, we consider the relaxation due to the CSA interaction for spins $j$ and $k$.
The relaxation superoperator due to the chemical shift anisotropy is diveide into two part, antisymmetric chemical shift anisotropy~(aCSA) and symmetric chemical shift anisotropy~(sCSA), and is written as:
\begin{eqnarray}
    \Gamma_{\mathrm{CSA}}=\sum_{r,s}\left(\Gamma^{r,s}_{\mathrm{aCSA}}+\Gamma^{r,s}_{\mathrm{sCSA}}\right), \quad
    r,s\in\{j,k\}.
\end{eqnarray}

\subsubsection{Antisymmetric chemical shift anisotropy}
The relaxation superoperator due to the antisymmetric part of the chemical shift interaction between spin $r$ and $s$ is described using Eq.~(\ref{eq:Spherical_representation_Hamiltonian}), Eq.~(\ref{eq:relax_master_equation}) and the values shown in Table~\ref{tb:Spherical_representation_Hamiltonian} as follows:
\begin{eqnarray}
    \Gamma^{r,s}_{\mathrm{aCSA}}
    &=&-\sum^{1}_{m,m^{\prime}=-1}\sum^{1}_{n,n^{\prime}=-1}(-1)^m  
    \tilde{T}^{r,\mathrm{aCSA}}_{(1,-m)} \tilde{T}^{s,\mathrm{aCSA}}_{(1,m^{\prime})} \nonumber \\
    &\times& \int_0^{\infty}
    \overline{A_{(1,n)}^{r,\mathrm{aCSA},\mathrm{M}}(0)
    D_{(n,m)}^{1}(\Omega^{r}_{\mathrm{ML}}(0))
    A_{(1,n')}^{s,\mathrm{aCSA},\mathrm{M},\ast}(-\tau)
    D_{(n',m')}^{1,\ast}(\Omega^{s}_{\mathrm{ML}}(-\tau))}
    \exp[-im^{\prime}\omega\tau] d\tau \nonumber \\
    &\simeq&-\frac{1}{3} 
    \left(\sum^{1}_{n=-1}
    A_{(1,n)}^{r,\mathrm{aCSA},\mathrm{M}}(0) A_{(1,n)}^{s,\mathrm{aCSA},\mathrm{M},\ast}(0)
    \right)
    \tau_{1}^{jk}
    \sum^{1}_{m=-1} (-1)^m  
    \tilde{T}^{r,\mathrm{aCSA}}_{(1,-m)} \tilde{T}^{s,\mathrm{aCSA}}_{(1,m)}.
\end{eqnarray}
Here, we assume that the correlation function decays with time in an exponential way,
use Wigner matrices property,
and $\tau_{1}$ is the rank-1 rotational correlation time of the molecule defined as
\begin{align}
    \tau_{1}^{jk} &= \int_0^{\infty}\overline{\cos(\theta^{jk}(\tau))}d\tau.
\end{align}
Here $\theta^{jk}(\tau)$ is the angle between the vector connecting the nuclear spin $j$ and $k$ axis at time $t=0$ and $t=\tau$.
The relaxation rate of the longitudinal magnetization is then calculated as
\begin{eqnarray}
    R_{\mathrm{1}}^{\mathrm{aCSA}}&=&
    -\frac{1}{2}
    \mathrm{Tr}[(I_{jz}+I_{kz})
    \sum_{r,s}\Gamma^{r,s}_{\mathrm{aCSA}}(I_{jz}+I_{kz})] \nonumber \\
    &=&
    \frac{\tau_{1}^{jk}}{3}B_0^2
    \left(||\sigma^{j,\mathrm{aCSA,M}}||^2+||\sigma^{k,\mathrm{aCSA,M}}||^2\right)
    \label{eq:R1_aCSA}
\end{eqnarray}
and the singlet state relaxation rate becomes
\begin{eqnarray}
    R_{\mathrm{S}}^{\mathrm{aCSA}}&=&- 
    \frac{4}{3} \mathrm{Tr}\left[(\mathbf{I}_j\cdot\mathbf{I}_k)
    \Gamma^{r,s}_{\mathrm{aCSA}}
    (\mathbf{I}_j\cdot\mathbf{I}_k)\right] \nonumber \\
    &=& 
    \frac{2}{9} \tau_1^{jk}
    B_0^2 \gamma_j^2
    ||\Delta\sigma^{\mathrm{aCSA}}||^2,
    \label{eq:RS_aCSA}
\end{eqnarray}
where $\Delta\sigma^{\mathrm{aCSA,M}}=\sigma^{j,\mathrm{aCSA,M}}-\sigma^{k,\mathrm{aCSA,M}}$.
$||\cdot||$ denotes the Frobenius norm. 

\subsubsection{Symmetric chemical shift anisotropy}
The relaxation superoperator due to the antisymmetric chemical shift interaction is described using Eq.~(\ref{eq:Spherical_representation_Hamiltonian}), Eq.~(\ref{eq:relax_master_equation}) and the values in Table~\ref{tb:Spherical_representation_Hamiltonian} as follows:
\begin{eqnarray}
    \Gamma^{r,s}_{\mathrm{sCSA}}
    &=&-\sum^{2}_{m,m^{\prime}=-2}\sum^{2}_{n,n^{\prime}=-2}(-1)^m  
    \tilde{T}^{r,\mathrm{sCSA}}_{(2,-m)} \tilde{T}^{s,\mathrm{sCSA}}_{(2,m^{\prime})} \nonumber \\
    &\times& \int_0^{\infty}
    \overline{A_{(2,n)}^{r,\mathrm{sCSA},\mathrm{P}}(0)
    D_{(n,m)}^{2}(\Omega^{r}_{\mathrm{PL}}(0))
    A_{(2,n')}^{s,\mathrm{sCSA},\mathrm{P},\ast}(-\tau)
    D_{(n',m')}^{2,\ast}(\Omega^{s}_{\mathrm{PL}}(-\tau))}
    \exp[-im^{\prime}\omega\tau] d\tau \nonumber \\
    &\simeq&-\frac{1}{5} 
    \left(\sum^{2}_{n=-2}
    A_{(2,n)}^{r,\mathrm{sCSA},\mathrm{P}}(0) 
    A_{(2,n)}^{s,\mathrm{sCSA},\mathrm{P},\ast}(0)
    \right)
    \tau_{2}^{jk}
    \sum^{2}_{m=-2} (-1)^m  
    \tilde{T}^{r,\mathrm{sCSA}}_{(2,-m)} \tilde{T}^{s,\mathrm{sCSA}}_{(2,m)}.
\end{eqnarray}
Here, we assume that the correlation function decays with time in an exponential way,
use Wigner matrices property,
and $\tau_{2}$ is the rank-2 rotational correlation time of the molecule defined as
\begin{align}
    \tau_{2}^{jk} &=\frac{1}{2}\int_0^{\infty}\overline{3\cos^2(\theta^{jk}(\tau))-1}d\tau.
\end{align}

The relaxation rate is then calculated as
\begin{eqnarray}
    R_{\mathrm{1}}^{\mathrm{sCSA}}&=&
    -\frac{1}{2}
    \mathrm{Tr}[(I_{jz}+I_{kz})
    \sum_{r,s}\Gamma^{r,s}_{\mathrm{sCSA}}(I_{jz}+I_{kz})] \nonumber \\
    &=& 
    \frac{1}{45}B_0^2\tau_{2}^{jk}
    \left((\delta^j)^2(3+(\eta^j)^2)+(\delta^k)^2(3+(\eta^k)^2) \right) \nonumber \\
    \label{eq:R1_sCSA}
\end{eqnarray}
and the singlet state relaxation rate due to the dipolar interaction becomes
\begin{eqnarray}
    R_{\mathrm{S}}^{\mathrm{sCSA}}&=&- 
    \frac{4}{3} \mathrm{Tr}\left[(\mathbf{I}_j\cdot\mathbf{I}_k)
    \Gamma^{r,s}_{\mathrm{sCSA}}
    (\mathbf{I}_j\cdot\mathbf{I}_k)\right] \nonumber \\
    &=& 
    \frac{2}{9}\tau_{2}^{jk}B_0^2\gamma_j^2
    ||\Delta\sigma^{jk,\mathrm{sCSA,M}}||^2,
    \label{eq:RS_sCSA}
\end{eqnarray}
where $\Delta\sigma^{jk,\mathrm{sCSA,M}}=\sigma^{j,\mathrm{sCSA,M}}-\sigma^{k,\mathrm{sCSA,M}}$.

\subsubsection{Spin--Rotation interaction}

As for the case of CSA, we consider the auto- and the cross-correlation terms of the SR interaction.
All relaxation superoperators for the spin-rotation interaction are written as
\begin{eqnarray}
    \Gamma_{\mathrm{SR}}=\sum_{r,s}\Gamma^{r,s}_{\mathrm{SR}}, \quad
    r,s\in\{j,k\}, \nonumber
\end{eqnarray}
\begin{align}
    \Gamma^{r,s}_{\mathrm{SR}} 
    &=-\sum_{m,m^{\prime}=-1}^{1} \sum_{p,p^{\prime}=-1}^{1} (-1)^{m+m^{\prime}}\nonumber \\
    &\times \int_{0}^{\infty} \overline{A_{1p}^{r,\mathrm{SR},\mathrm{M}}(0) A_{1p^{\prime}}^{s,\mathrm{SR},\mathrm{M},\ast}(-\tau)
    D_{pm}^1(\Omega_{r,\mathrm{ML}}(0)) D_{p^{\prime}m^{\prime}}^{1,\ast}(\Omega^{s}_{\mathrm{ML}}(-\tau))} \tilde{T}_{-m}^{r,\mathrm{SR}} \tilde{T}_{-m'}^{s,\mathrm{SR},\dagger} d\tau \nonumber \\
    &=-\frac{1}{3}\sum_{m=-1}^{1}(-1)^{m}T_{-m}^{r,\mathrm{SR}} T_{m}^{s,\mathrm{SR}}
    \sum_{p=-1}^{1} 
    \int_{0}^{\infty} 
    \overline{A_{1p}^{r,\mathrm{SR},\mathrm{M}}(0) A_{1p}^{s,\mathrm{SR},\mathrm{M},\ast}(-\tau)}
    \exp[-im\omega \tau] d\tau \nonumber \\
    &=-\frac{1}{3}\sum_{m=-1}^{1}(-1)^{m}T_{-m}^{r,\mathrm{SR}} T_{m}^{s,\mathrm{SR}}
    \sum_{\alpha\in\{x,y,z\}}
    \left( C^{\mathrm{M}}_{x\alpha,r}C^{\mathrm{M}}_{x\alpha,s} +C^{\mathrm{M}}_{y\alpha,r}C^{\mathrm{M}}_{y\alpha,s}+C^{\mathrm{M}}_{z\alpha,r}C^{\mathrm{M}}_{z\alpha,s}\right)
    \frac{\mathrm{I}_{\alpha}^2}{6\hbar^2\tau_2^{jk}}.
\end{align}
In the last equation, we assume the FML regime and use Hubbard's rule~\cite{hubbard1963theory} and rank-2 correlation time. 

Consequently, the longitudinal relaxation rate due to the spin-rotation interaction becomes
\begin{eqnarray}
    R_{\mathrm{1}}^{\mathrm{SR}}&=&-\frac{1}{2}\sum_{r,s}\mathrm{Tr}[(I_{jz}+I_{kz})\Gamma^{r,s}_{\mathrm{SR}}(I_{jz}+I_{kz})] \nonumber \\
    &=&\frac{1}{9\hbar^2 \tau_2^{jk}}\sum_{r=j,k}\sum_{\alpha}\left( (C^{\mathrm{M},r}_{x\alpha})^2 +(C^{\mathrm{M},r}_{y\alpha})^2 + (C^{\mathrm{M},r}_{z\alpha})^2\right)
    \mathrm{I}_{\alpha}^2
    \label{eq:R1_SR}
\end{eqnarray}
and the singlet state relaxation rate due to the spin-rotation interaction becomes
\begin{eqnarray}
    R_{\mathrm{S}}^{\mathrm{SR}}&=&
    -\frac{4}{3}\sum_{r,s}\mathrm{Tr}[T_{00}^{jk}\Gamma^{r,s}_{\mathrm{SR}}T_{00}^{jk}] \nonumber \\
    &=&
    \frac{1}{9\hbar^2 \tau_2^{jk}}\sum_{\alpha}\sum_{\beta}
    \left( |\Delta C^{\mathrm{M}}_{\beta\alpha}|^2\right)
    \mathrm{I}_{\alpha}^2,
    \nonumber \\
    \label{eq:RS_SR}
\end{eqnarray}
where $\Delta C^{\mathrm{M}}_{\beta\alpha}=C^{\mathrm{M},j}_{\beta\alpha}-C^{\mathrm{M},k}_{\beta\alpha}$.

\subsection{Computational details}

\begin{figure}
    \centering
    \includegraphics[width=170mm]{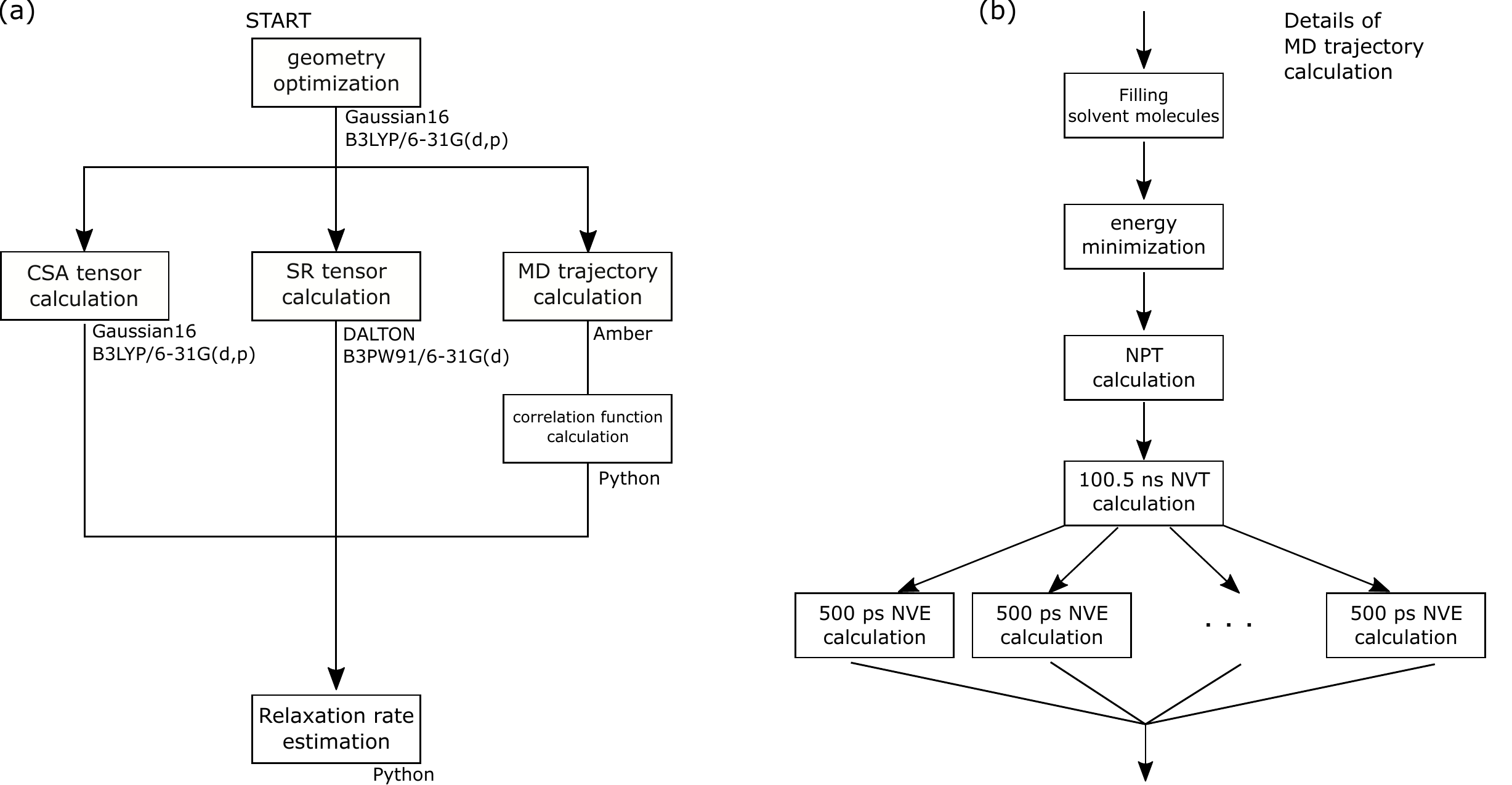}
    \caption{Flowchart for relaxation rate calculation using MD trajectories and QC tensor calculations.}
    \label{fig:Flowchart}
\end{figure}

% この文章はcomputatinal detailsに回す。
All processings were performed on a custom-built workstation running CentOS Linux 7 with an AMD EPYC 7742 64-Core Processor and NVIDIA A100 GPUs.
Flowchart for the calculation of the relaxation rates is shown in Fig.~\ref{fig:Flowchart}.

\subsection{Molecular dynamics method}
The initial molecule geometries were designed using the Avogadro 1.1.1~\cite{Hanwell2012} molecule editor and relaxed to the lowest-energy geometry through a steepest-decent algorithm.
Then, the molecule geometries were optimized using Gaussian~16~\cite{g16} at the B3LYP/6-31G(d) level of theory.  
MD method was performed with the SANDER or pmemd.cuda module of the AMBER program package~\cite{gotz2012routine,salomon2013routine,le2013spfp}.
Some of the target molecules are in the anionic state depending on the solvent.

\begin{figure}
    \includegraphics[width=100mm]{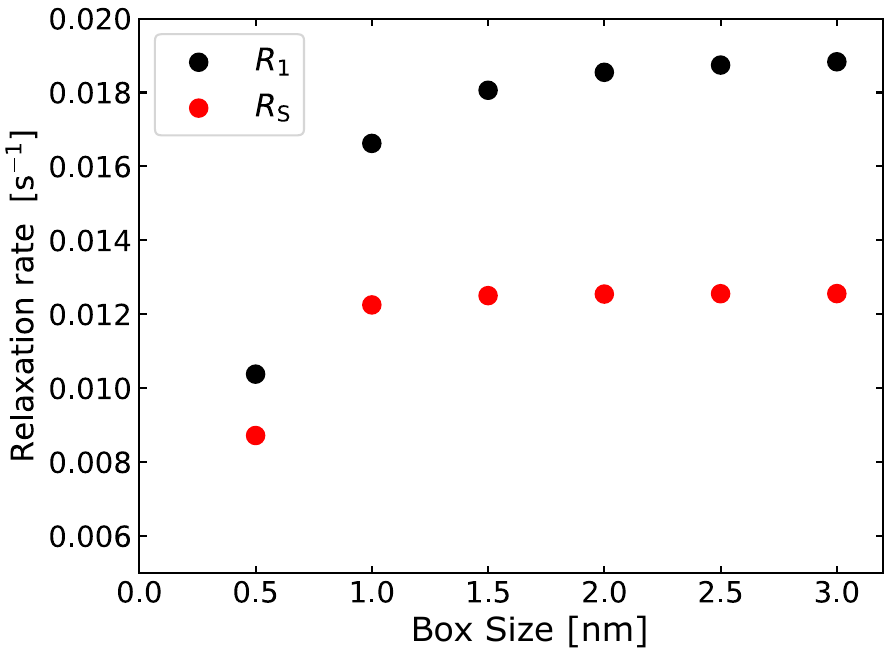}
    \caption{The intermolecular dipolar relaxation dependence of the simulation box size.}
    \label{fig:vd_dependence}
\end{figure}

The target molecules were solvated with several solvates including water, methanol, acetone, and dimethyl sulfoxide (DMSO).
We used TIP4P~\cite{jorgensen1983comparison} for water and the meoh model provided by the AMBER~(MEOHBOX) program. For the other molecules, we employed the General Amber Force Field (GAFF).
Although these solvents were not substituted with deuterium, the gyromagnetic ratio for deuterium was used in the relaxation rate calculation instead of that for proton.
These solvates were put in a 3.0~nm cube using the solvatebox command of tLeap, and sodium ion was added to make the entire system neutral.
The box size of 3.0~nm was determined by the simulation results of the intermolecular dipolar relaxation for several box sizes.
We confirmed that the calculated intermolecular dipolar relaxation rate increased with the box size, and this relaxation rate is saturated after 3~nm as shown in Fig.~\ref{fig:vd_dependence}.
In this calculation, the aromatic protons in $p$-aminobenzoic acid (PABA) dissolved in TIP4P water were used.

The molecular dynamics simulations were performed in the following steps: (i) the initial energy minimization of the system was run; (ii) NPT calculation over 1~ns was run to determine the volume of the system; (iii) then, NVT calculation over 100.5~ns was run; (iv) the snap shots were extracted every 500~ps from the NVT calculation results; (v) 200 NVE trajectories over 500~ps were run using these 200 extracted geometries and velocities. 
In the molecular dynamics simulations, we used the Langevin Thermostat.
The force integration and coordinate/velocity information were updated every 0.5~fs using Beeman’s algorithm, with coordinates output every 0.5~ps. 
Finally, the correlation times and the averaged distances were calculated using the 200 NVE calculation results.
In these calculations, we used the Langevin Thermostat.
Finally, the correlation functions were estimated from the average of the 200 NVE trajectories.

% 計算の流れ参考用
% antechamber 19.0  量子化学計算のアウトプットをMD用にする、gaussianの計算結果から対象分子の力場を計算
% tleap             MD計算を行う系の構築 gaffの意味は多分、gaussian計算以外の動的なデータの取得 イオン、溶媒を入れる。
% amber 20 sander minimize　minimize.out
% amber 20 sander NPT md.out
% cpptraj v4.14.0
% amber 18 PMEMD 2018 NVT
% NPT 1ns
% NVT 100.5ns
% NVE 100ns

\subsection{Quantum Chemistry Calculation}
The quantum chemical calculations for the chemical shielding tensor, the moment of inertia of the molecule, and the spin-rotation tensor were performed using Gaussian 16 Revision A.03~\cite{ogliaro2016gaussian} and DALTON 2018.0 software~\cite{aidas2014d}.
The geometry optimizations were performed using Gaussian at the B3LYP/6-31G(d,p) level of theory, where the initial geometries of the molecules were designed using the Avogadro 1.1.1~\cite{Hanwell2012} molecule editor and relaxed to the lowest-energy geometry through a steepest-decent algorithm.
The chemical shielding tensors were computed using Gaussian at the B3LYP/6-31G(d,p) level of theory. In the calculation using the Gaussian program suite, we used the polarizable continuum model to consider the effects of the solvents. The spin-rotation tensors and the moment of inertia were computed using the DALTON at the B3PW91/6-31G(d) level of theory.

\subsection{Relaxation rates}
The detailed values of the calculated relaxation rates are shown in Table~\ref{tb:Calculation_results_T1} and Table~\ref{tb:Calculation_results_TLLS}.

\begin{table}[h]
    \centering
    \caption{Experimental and calculation values of longitudinal relaxation time.} 
    \begin{tabular}{ccccccccc} \hline \hline
                                  &                  &  \multicolumn{4}{c}{$1/T_1$ (10$^{-2}\times$s$^{-1}$)}    & & \multicolumn{2}{c}{$T_1$ (s)} \\ \cline{3-6}\cline{8-9}
    molecule                       & solvent &DDjk\quad\quad          &inDD\quad\quad          &exDD\quad\quad  & CSA     &&Calc.\quad\quad& ~Exp.~ \\ \hline
    PABA~\cite{Pileio06}           &   water &29.6$\,\pm\,$4   \quad\quad&1.55$\,\pm\,$0.17\quad\quad& 1.93$\,\pm\,$0.04\quad\quad & 1.16 && 2.92$\,\pm\,$0.36\quad\quad & 5.15 \\ 
    PCBA~\cite{miyanishi2020long}  &   water &27.4$\,\pm\,$3.7 \quad\quad&1.31$\,\pm\,$0.14\quad\quad& 2.1 $\,\pm\,$0.05\quad\quad & 1.25 && 3.12$\,\pm\,$0.38\quad\quad &  5.3 \\
    PHBA~\cite{Pileio06}           &   water &37.8$\,\pm\,$4.3 \quad\quad&2.19$\,\pm\,$0.17\quad\quad& 2.47$\,\pm\,$0.06\quad\quad & 1.16 && 2.29$\,\pm\,$0.24\quad\quad & 4.71 \\
    Cl-TC~\cite{kiryutin2019proton}& acetone &1.56$\,\pm\,$0.63\quad\quad&0.11$\,\pm\,$0.02\quad\quad& 0.13          \quad\quad & 0.02 && 55$\,\pm\,$19.6\quad\quad & 47 \\
    Cl-TC~\cite{kiryutin2019proton}& DMSO    &6.04$\,\pm\,$1.65\quad\quad&0.16$\,\pm\,$0.02\quad\quad& 0.55$\,\pm\,$0.02\quad\quad & 0.11 && 14.6$\,\pm\,$3.61\quad\quad & 11.9 \\
    Cl-TC~\cite{kiryutin2019proton}& methanol&3.91$\,\pm\,$0.8\quad\quad&0.05$\,\pm\,$0.02\quad\quad& 0.26          \quad\quad & 0.03 && 23.5 $\,\pm\,$4.56\quad\quad & 32.8\\
    Cl-TC~\cite{kiryutin2019proton}& water   &8.82$\,\pm\,$2.2 \quad\quad&0.15$\,\pm\,$0.02\quad\quad& 1.59$\,\pm\,$0.04\quad\quad & 0.49 && 9.05 $\,\pm\,$1.81\quad\quad & 18.7\\ 
    \hline \hline
    \end{tabular}
  \label{tb:Calculation_results_T1}
\end{table}

\begin{table}[h]
    \centering
    \caption{Experimental and calculation value of relaxation time for long-lived state.}
    \begin{tabular}{cccccccc} \hline \hline
                                  &                 &  \multicolumn{3}{c}{$1/T_{\mathrm{S}}$ (10$^{-3}\times$s$^{-1}$)}    & & \multicolumn{2}{c}{$T_{\mathrm{S}}$ (s)} \\\cline{3-5}\cline{7-8}
        ~ molecule~                &solvent &inDD           \quad\quad&exDD~         \quad\quad&CSA           \quad\quad&&Calc.      \quad\quad& Exp. \\ \hline
    PABA~\cite{Pileio06}           &water   &16.2 $\,\pm\,$1.23\quad\quad&12.9$\,\pm\,$0.7 \quad\quad&7.66$\,\pm\,$0.02\quad\quad&&27.2$\,\pm\,$1.4\quad\quad& 20 \\ 
    PCBA~\cite{Pileio06}           &water   &10.7 $\,\pm\,$0.97\quad\quad&16.3$\,\pm\,$0.68\quad\quad&9.08$\,\pm\,$0.03\quad\quad&&27.7$\,\pm\,$1.3\quad\quad& 15   \\ 
    PHBA~\cite{miyanishi2020long}  &water   &19.5 $\,\pm\,$1.64\quad\quad&19.6$\,\pm\,$0.9 \quad\quad&6.23$\,\pm\,$0.03\quad\quad&&22.1$\,\pm\,$1.2\quad\quad& 21.7   \\ 
    Cl-TC~\cite{kiryutin2019proton}&acetone &0.94 $\,\pm\,$0.25\quad\quad&0.59$\,\pm\,$0.06\quad\quad&0.09          \quad\quad&&616 $\,\pm\,$120\quad\quad& 183  \\
    Cl-TC~\cite{kiryutin2019proton}&DMSO    &1.64 $\,\pm\,$0.31\quad\quad&2.59$\,\pm\,$0.26\quad\quad&0.45          \quad\quad&&214 $\,\pm\,$25.9\quad\quad& 134  \\
    Cl-TC~\cite{kiryutin2019proton}&methanol&0.58 $\,\pm\,$0.29\quad\quad&1.1 $\,\pm\,$0.11\quad\quad&0.13          \quad\quad&&550 $\,\pm\,$121\quad\quad& 236  \\
    Cl-TC~\cite{kiryutin2019proton}&water   &2.07 $\,\pm\,$0.28\quad\quad&9.11$\,\pm\,$0.11\quad\quad&1.45          \quad\quad&&79.1$\,\pm\,$2.5\quad\quad& 132 \\
    \hline \hline
    \end{tabular}
  \label{tb:Calculation_results_TLLS}
\end{table}

\end{document}